\documentclass[journal=jpcc,manuscript=article,layout=onecolumn]{achemso}

\usepackage[version=3]{mhchem} 
\usepackage{chemarr}

\usepackage{textcomp, amssymb, mathrsfs}

\usepackage{epstopdf}
\usepackage{soul,color}

\usepackage{circuitikz,tikz}

\author{Anis Allagui}
\email{aallagui@sharjah.ac.ae}
\affiliation[UOS]{Dept. of Sustainable and Renewable Energy Engineering, University of Sharjah, Sharjah, P.O. Box 27272, United Arab Emirates}
\altaffiliation{Center for Advanced Materials Research, Research Institute of Sciences and Engineering, University of Sharjah, Sharjah,, P.O. Box 27272,  United Arab Emirates}
\affiliation[FIU]{Dept. of Mechanical and Materials Engineering, Florida International University, Miami, FL33174, United States}

\author{Hachemi Benaoum} 
\affiliation[UOS]{Dept. of Applied Physics and Astronomy, University of Sharjah, PO Box 27272, Sharjah, United Arab Emirates}

\author{Chunlei Wang} 
\affiliation[FIU]{Dept. of Mechanical and Materials Engineering, Florida International University, Miami, FL33174, United States}

\title{Deformed Butler-Volmer Models for Convex Semilogarithmic Current-Overpotential Profiles of Li-ion Batteries}

\begin{document}

\begin{abstract}

The Butler-Volmer (BV) equation links the current flux crossing an electrochemical interface to the electric potential drop across it with the assumption of Arrhenius kinetics and the Boltzmann factor. Applying the semilogarithmic  Tafel   analysis in which the logarithm of current is plotted vs. the overpotential one expects straight lines from which the fundamental reaction rate of the kinetic process can be computed. However, some Li-ion battery data, which is the focus here,  show nonlinear convex profiles that cannot be adequately fitted with the standard BV  model. We propose instead two deformed BV models for the analysis of such types of   behaviors constructed from the superposition of cells exhibiting only local equilibrium and thus giving rise to the power-law   $q$-exponential and $\kappa$-exponential functions. Non-Boltzmann distributions have been successfully employed for the modeling of a wide spectrum of physical systems in nonequilibrium situations, but not yet for batteries. We verify the validity of the deformed BV models on experimental data obtained from \ce{LiFePO4} and \ce{Li}-\ce{O2} batteries.

\end{abstract}

\section{Introduction}
\label{introduction} 
Chemical-to-electrical or electrical-to-chemical energy conversion in electrochemical energy devices and systems such as batteries and fuel cells or in electrodeposition and electrolysis require the presence of two phases in contact with each other.  Generally there is an electron-conducting phase but ionic insulator on the one hand, and an ionic conductor but electric insulator on the other hand. The phases can be solid, liquid or gas. At the inter-phase boundary or interface, which should be thought of as a physical and not a mathematical plane, chemical reactions accompanied with charge carriers transfer from one phase to another take place.  Understanding and properly modeling the measurements carried out on such systems is important for many applications including renewable energy technologies with energy storage, electric vehicles, smart grids, and corrosion prevention  \cite{bai2014charge}. In particular, it is important and useful to describe the system-level relationship between the electrical current flowing through an electrode, or charge dynamics with time, and the potential difference between the electrode itself and a point in the bulk electrolyte. 
 Several approaches have been proposed to model kinetic rate behavior.   The macroscopic phenomenological Butler–Volmer (BV) model is the de facto  mathematical model used for describing   the simultaneous anodic  reaction  (oxidation) and cathodic  reaction (reduction) on the same electrode surface \cite{dreyer2016new}.  
 Microscopic theories, including Marcus, Marcus-Hush-Chidsey (MHC)   and their extensions are also powerful  in describing electron transfer kinetics in both directions with physically-tractable quantities \cite{Marcus, Chidsey}. However, the main limitation for  the widespread use of Marcus-type theories is that the rate has no simple closed-form expression. It is  defined in terms on indefinite integral over Fermi-Dirac distribution function for which different computational algorithms have been proposed with different levels of accuracy  and computational costs \cite{MHS-DOS}. Our focus here is on the BV model. 
 
 The BV equation can be derived from different approaches   including  the kinetic law of mass action \cite{gutman2007empiricism}, non-equilibrium thermodynamics \cite{dreyer2016new},   principle of thermal activation process \cite{yang2016generalized}, and also from first principles \cite{fletcher2009tafel}, but here we present the derivation from  transition state theory \cite{vetter2013electrochemical}.  From Arrhenius kinetics, it is observed that the natural logarithm of the reaction rate $k$ and the reciprocal of the absolute temperature $T$ are linearly related according to:
\begin{equation}
\frac{\partial \ln(k)}{\partial(1/T)} = -\frac{E_a}{R}
\end{equation}
where  $R$ is the universal gas constant and $E_a$ is an experimental/phenomenological activation energy \cite{gutman2007empiricism}, which can be thought of as the energy necessary to overcome a certain energy barrier for particles  to transition from the well of reactants to the well of products, and thus   for the reaction to proceed. Arrhenius' suggestion that there is a  transition state intermediate between reactants and products was central to the development of transition state theory \cite{pollak2005reaction}. 
 Let us consider the case of a single-step charge transfer redox reaction  of the form:
\begin{equation}
\ce{Ox} + n \ce{e}^- \rightleftharpoons \ce{Red}
\label{eq1}
\end{equation}
The forward or oxidation ($\ce{Ox} + n \ce{e}^- \rightarrow \ce{Red}$) and backward or reduction ($ \ce{Red} \rightarrow \ce{Ox} + n \ce{e}^- $) charge transfer reactions can be described by the charge transfer reaction rates $k_f$ and $k_b$. Their associate forward and backward current densities (per unit surface) crossing the interface 
 are taken to be proportional to the surface concentrations $C_{\text{Ox}}$ and $C_{\text{Red}}$, and can be written as:
 \begin{eqnarray}
i_f = n F k_f C_{\text{Ox}} \\
i_b =  n F k_b C_{\text{Red}}
\end{eqnarray}
 where $F$ is the Faraday constant. 
 Thus the net current flowing through the electrode is given by the difference \cite{gutman2007empiricism}:
 \begin{equation}
i=i_f - i_b = n F ( k_f C_{\text{Ox}} - k_b C_{\text{Red}} )
\label{eqi}
\end{equation} 
 Now by incorporating the expressions for the potential and temperature-dependence of the forward and backward charge transfer reaction rates, which are assumed to follow Arrhenius profiles (with  $E_a$ taken as a linear function of the potential $\psi$) as:
 \begin{eqnarray}
k_f = k_f^0  \exp \left( \frac{-\alpha_{\text{Red}} \psi}{RT/F} \right) 
\label{eqkf} \\
k_b = k_b^0 \exp \left( \frac{\alpha_{\text{Ox}} \psi}{RT/F} \right)
\label{eqkb}
\end{eqnarray} 
Eq.\;\ref{eqi}  turns to be:
\begin{equation}
{i}={n F} \left\{  k_f^0 C_{\ce{Ox}} \exp \left( \frac{-\alpha_{\text{Red}} \psi}{RT/F} \right) -  k_b^0 C_{\ce{Red}} \exp \left( \frac{\alpha_{\text{Ox}} \psi}{RT/F} \right) \right\}
\label{eqBV}
\end{equation}
 This equation  is known as the kinetic  BV   equation for the current-potential relationship with pure charge transfer overpotential. 
Here,  the dimensionless parameters  
 $\alpha_{\ce{Red}}$  and $\alpha_{\ce{Ox}}$  (taking values between 0 and 1 with $\alpha_{\ce{Red}}+\alpha_{\ce{Ox}}=1$) denote  
 the  transfer or symmetry factors  associated with the  oxidation and reduction reactions,  respectively, 
 or qualitatively a measure of the "position" of the transition state \cite{li2018tafel},  
  $RT/F=V_{th}$ is the thermal voltage, and  $\psi$ is the potential of the electrode through which current flows, which is different from the equilibrium potential $\psi_0$ established when no current passes through the electrode.
   The difference $\eta=\psi-\psi_{0}$  is known as the overpotential. 
 Eq.\;\ref{eqBV} can also be expressed in terms of the current exchange density $i_0$  (i.e. when $i_f=i_b=i_0$ which takes place at the equilibrium potential $\psi_{0}$) such that: 
\begin{equation}
{i}=i_0 \left\{    \exp \left( \frac{-\alpha_{\text{Red}} \eta}{RT/F} \right) -    \exp \left( \frac{\alpha_{\text{Ox}} \eta}{RT/F} \right) \right\}
\label{eqBVio}
\end{equation}  
  Note that with the use of the dimensionless overpotential scaled to the thermal voltage, $\eta^*=\eta / V_{th}$, the dimensionless current $i^*=i/i_0$, and $\alpha_{\text{Red}}= \alpha$, Eq.\;\ref{eqBVio} can be rewritten as \cite{bai2014charge}:
 \begin{equation}
i^* = \exp \left[-\alpha \eta^*\right] - \exp \left[ (1-\alpha)\eta^* \right]
\label{eq10}
\end{equation}
 For the particular case of $\alpha=0.5$, which is commonly used for battery modeling, the expression for $i^*$ simplifies to:
 \begin{equation}
i^* = 2 \sinh (- \eta^*/2)
\end{equation}
The Tafel technique of plotting  $\ln (i^*)$ versus $\eta^*$ gives a straight line of slope $-\alpha$ for $\eta^*<0$ and $(1-\alpha)$ for $\eta^*>0$ from which the charge transfer rates can be estimated \cite{bai2014charge, banham2009pt, sankarasubramanian2017elucidating, soderberg2006impact}. 

 The validity of the BV model is based on the assumption that the concentration of the reacting species are independent of the current density and the potential, and consequently only pure charge transfer overpotential is involved \cite{vetter2013electrochemical}. 
 It is also commonly assumed that the transfer coefficient 
 is independent or a weak function  of the applied potential and can be considered as constant \cite{li2018tafel}. 
 The  surface of the electrode is considered flat and   stress-free \cite{yang2016generalized}. 
Furthermore, within the BV framework  an exponential Boltzmann factor for the reaction rate dependence on the temperature is considered. The Boltzmann factor, which is essentially a comparison between the energy of the molecules and the energy of the barrier when the system is in thermodynamic equilibrium and characterized by a certain temperature,  assumes that particles are totally independent, non-interacting and obey the laws of ideal gases \cite{allagui2021gouy}. It also assumes that elementary volumes of the system are equiprobable. These   assumptions are the basis for  Boltzmann-Gibbs (BG) statistical mechanics in which the exponential and Gaussian distributions are those  that maximize the BG entropy by virtue of the the Central Limit Theorem (CLT), and ensure the equilibrium state.

However, there are several instances where the semilogarithmic Tafel analysis of $\ln (i^*)$ versus $\eta^*$ does not result in linear profiles, but rather   curved plots which is the motivation for this work \cite{bai2014charge, boyle2020transient, munakata2012evaluation, viswanathan2013li, sankarasubramanian2017elucidating, soderberg2006impact}. Focusing on battery materials and devices, curved Tafel plots have been reported for instance by Munakata et al.  \cite{munakata2012evaluation} in experiments conducted on  single (porous) particles of \ce{LiFePO4}, which is a widely used cathode material for large-scale batteries. The fitting with the BV model with the symmetry factor $\alpha=0.5$ was reasonable enough for a small portion of the voltage-current data only  \cite{munakata2012evaluation}. 
Viswanathan et al. \cite{viswanathan2013li}  reported also highly nonlinear Tafel plots for the discharge of  nonaqueous Li-air (or \ce{Li}-\ce{O2}) battery. The charging data were less unusual by showing slight nonlinearities in the profiles. 
 Generally speaking, reacting systems we are interested in are actually   away from equilibrium and transition from one metastable state to a neighboring state of metastable equilibrium in response to external stimuli \cite{yin2014collision}. Thus, the assumption of thermodynamic equilibrium is not always appropriate, and the statistics may not necessarily  follow   BG  statistics \cite{pollak2005reaction,  yin2014collision, allagui2021gouy}. 
 In fact, for many complex systems at off-equilibrium conditions  it is often observed that power-law distributions are most common as it is the case for example  with the dissolution reaction of magnesium (or aluminum) in aqueous cupric chloride solution \cite{claycomb2004power}. Magnesium (or aluminum) dissolves to form \ce{MgCl2} (or \ce{AlCl3}) and copper precipitates at local reaction rates that can be affected by concentration fluctuations, pitting dissolution and the formation of the Cu layer, which tends to inhibit the reaction itself. Furthermore, the breakdown of the Cu layer because of liberation of hydrogen gas, convective turbulence near the reactive surface, and erosion of underlying metal, increase the local reaction rate at the freshly exposed   surfaces.  As a result, one observes fluctuations in current and voltage found to follow power-law behavior.  \cite{claycomb2004power}. 
  Other systems that exhibit  power-law statistics  are for example the power grid frequency fluctuations   \cite{schafer2018non}, epidemiology and spreading dynamics of diseases  \cite{kaniadakis2020kappa}, and atomic packing in metallic glasses \cite{zhang2019power} to name a few.

The purpose of this contribution is to formulate and study a generalized BV model by incorporating the power-law 
(i)  $q$-exponential  function   based on Tsallis  nonextensive statistics \cite{tsallis1988possible, tsallis2019beyond} 
 and (ii) the $\kappa$-exponential based on Kaniadakis statistics \cite{kaniadakis2001non,kaniadakis2013theoretical}  
instead of the traditional  exponential   Boltzmann factor based on BG statistics.  
Such approaches have been proven  successful  for describing many  complex nonequilibrium  systems at stationary state that behave like the collective superposition of many subsystems, themselves they follow the BG statistics. 
 These systems usually involve long-range interactions, non-Markovian memory effects and anomalous diffusion for instance. The single Boltzmann factor employed in the BV kinetic model is recovered as a  limiting case. We note  that the generalization of reaction rate coefficients using the $q$-exponential structure instead of the standard Arrhenius exponential function has been proposed by Niven \cite{niven2006q}, Bagci \cite{baugci2007nonextensive},   Yin et al. \cite{yin2014power, yin2014rate, cangtao2014power} amongst others \cite{aquilanti2010temperature, quapp2010transition, silva2013uniform, aquilanti2017kinetics}, but to the best of our knowledge this is the first study on the extension of the BV model using such a framework for analyzing battery data.  Furthermore, we are not aware of any studies using the $\kappa$-exponential function to do so. 
 
 The rest of the manuscript is organized as follows. In Section\;\ref{theory} we will provide a brief summary of some important deformed functions (mostly deformed exponential functions) and their properties, and formulate the corresponding modified BV expressions. In Section\;\ref{simulation}  we analyze experimental Lithium batteries results compiled from the literature for which we compare fittings using   standard, and $q$- and $\kappa$-deformed BV equations.  

\section{Theory}
\label{theory}

We consider the simplified case model of an electrode/electrolyte system with single-step charge transfer redox reaction as described by reaction\;\ref{eq1}. The global equilibrium of the reacting system, which is driven away from equilibrium, is assumed to be influenced by fluctuations and stochastic events. 
 Such fluctuations can originate for instance from the effects of the non-uniformity of the electrode/electrolyte interfaces, porous and fractal structures, long-range interactions, irreversibilities and parasitic reactions, particle trapping and partial charge transfer, as well as local variations in thermodynamic parameters. 
    When modeling such stochastic dynamics, the question that comes first is how to approximate the noise distribution? This can be modeled on the one hand using non-Gaussian distributions when fluctuations are known to display heavy tails and skewness  such as in the form of Levy-stable distributions \cite{schafer2018non}, or on the other hand the underlying stochastic process can be interpreted as a superposition of multiple Gaussian distributions, leading to the framework of Beck and Cohen  superstatistics \cite{beck2004superstatistics, abe2007superstatistics, beck2003superstatistics, schafer2018non}, which is also able to explain heavy tails and skewness. 
     We consider here the latter with the assumption that the macroscopic electrode/electrolyte system is  made up of many subsystems that are temporarily in local equilibrium, but each has different statistics (e.g. standard deviation).  
     This can also be viewed from the angle that the process finds an equilibrium with an approximately Gaussian distribution determined by the current noise, and then after a lapse of time large enough compared with the intrinsic timescale of the system, the system finds another equilibrium also following an approximately Gaussian distribution but with different statistics  \cite{schafer2018non}. 
  In other words, we are considering the situation in which the total distribution of the  transfer reaction rate can be viewed as several aggregated Gaussian distributions, making it not only dependent on the potential and absolute temperature as it is the case in Eqs.\;\ref{eqkf} and \ref{eqkb}, but also on the extent and statistics  of   fluctuations superposed on the equilibrium.  The fluctuating variable could be for instance the inverse temperature $\beta=1/T$ (or in energy units $\beta=1/(k_B  T)$) or any other intensive quantity \cite{beck2003superstatistics, beck2004superstatistics}. 

\begin{figure}[t]
\begin{center}
\includegraphics[height=1.45in]{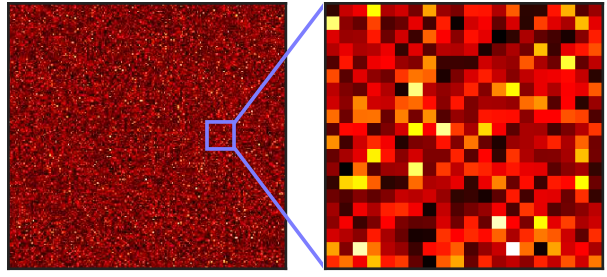}
\caption{Illustration  of a discretized electrode system into a number of spatial cells with different  values of inverse temperatures $\beta$ (derived in this example from a Gamma distribution   with unit scale and shape parameter 3). In the left side we show a low magnification of the electrode whereas in right side a zoomed area is depicted}
\label{fig0}
\end{center}
\end{figure} 

 In Fig.\;\ref{fig0} we show a schematic illustration of how one can imagine the inhomogeneous electrode system to look like when discretized into a number of  spatial cells with different values of $\beta$. Each cell is large enough so that it can be represented by a constant value of $\beta$, and thus a single Boltzmann factor is valid. In this regard, from Beck and Cohen \cite{beck2003superstatistics, beck2006stretched},  a generalized Boltzmann factor for the whole system can be written as the integral over all possible fluctuating inverse temperatures $\beta$ of Boltzmann factors $\exp({-\beta E})$ as:  
\begin{equation}
B(E) = \int\limits_0^{\infty} f(\beta) \exp({-\beta E})  \mathrm{d}\beta 
\label{eq:BE}
\end{equation}
 where $f(\beta)$ represents a normalized probability distribution function (PDF, with $\int_{-\infty}^{\infty} f(\beta)  \mathrm{d} \beta =1$) and provides a weight for the   distributions $\exp({-\beta E})$ \cite{beck2003superstatistics}.   $B(E)$ in Eq.\;\ref{eq:BE} is essentially the Laplace transform of the function $f(\beta)$ \cite{beck2006stretched}, and $f(\beta)$ is here to reshape the Boltzmann distribution into a generalized Boltzmann distribution by providing a statistics for the BG theory statistics, and thus \textit{superstatistics} of Beck and Cohen.
In other words, Eq.\;\ref{eq:BE} can be used to describe a macroscopic nonequilibrium system in a stationary state such as the case of electrified electrode/electrolyte system, but locally the system shall remain infinitely close to equilibrium for which the theory of equilibrium statistical mechanics holds \cite{ourabah2021fingerprints}.

\subsection{$q$-deformed BV model}
 
Considering the generalized Boltzmann factor given by Eq.\;\ref{eq:BE}, we now derive an extension to  the traditional Boltzmann exponential behavior depending on  the choice of the density function $ f(\beta)$.  
Particularly, it is known that the sum of $n$ independent exponentially-distributed variables of PDF equal to $a \exp(-a \beta)$ where $a>0$ has the discrete PDF \cite{bondesson2012generalized}:
\begin{equation}
f_n(\beta) = \frac{a^n}{\Gamma(n)} \beta^{n-1} \exp \left(- a \beta \right)
\label{eqg}
\end{equation}
By replacing $n$ in Eq.\;\ref{eqg} by any real positive number, we get the general (continuous) Gamma distribution \cite{lienhard1967physical}. 
 For the case of $n=1$ one retrieves the exponential distribution, but a number of other distributions can be obtained as special cases, such as the chi-square, Weibull,  hydrograph,  Rayleigh  or the Maxwell molecular velocity distributions \cite{lienhard1967physical}. This makes the Gamma distribution versatile enough to describe different types of statistics  \cite{beck2006stretched}.  
 We mention that a further generalization of the Gamma distribution (associated with a Bessel function for instance) can be written as \cite{sebastian2011generalized}:
 \begin{equation}
f_B(\beta) = \frac{a^\gamma}{\Gamma(\gamma) \exp(x/a)} \beta^{\gamma-1} \exp(-a \beta)\, _0 F_1 (; \gamma;  \beta x)
\label{eqSeb}
\end{equation}
where $\beta>0$, $\gamma>0$, $a>0$ and 
$_0 F_1 (; \gamma; \beta x) = \sum_0^{\infty} (  \beta x) ^k / (\gamma)_k k! $,  
$(\gamma)_k$ is the Pochhammer symbol 
 ($(\gamma)_k = \gamma(\gamma+1) \ldots (\gamma+k-1)$, $\gamma \neq 0$, $(\gamma)_0=1$).
It is clear that Eq.\;\ref{eqSeb} with  $x = 0$ reduces to  the  distribution: 
\begin{equation}
f_q(\beta) = \frac{1}{b \Gamma(c)} \left({\beta}/{b}\right)^{c-1} \exp \left(-{\beta}/{b} \right)
\label{eqgamma}
\end{equation}
in which we used Beck and Cohen's notations \cite{beck2003superstatistics}
 where $c$ and $b$ are positive parameters.  
 The integration over $\mathrm{d}\beta$  in Eq.\;\ref{eq:BE} with $f(\beta)$ given by  Eq.\;\ref{eqgamma} leads unambiguously  to the closed-form power-law function \cite{beck2003superstatistics}:
\begin{equation}
B(E) = (1+bE)^{-c} 
\label{eq:Tsallis}
\end{equation}
With the substitutions  $-1/(1-q)=c$ and  
$\beta_0= \int_0^{\infty} \beta f_q(\beta) \mathrm{d}\beta 
= bc$ being the average of the fluctuating $\beta$, the r.h.s of Eq.\;\ref{eq:Tsallis} is rewritten as \cite{beck2003superstatistics,tsallis1988possible}:
\begin{equation}
(1+bE)^{-c} = \left[ 1- (1-q)\beta_0 E \right]^{{1}/{(1-q)}} \equiv \exp_q({-\beta_0 E})
\label{eq:Tsallis1}
\end{equation} 
where $\exp_q(y)$ denotes  the  $q$-exponential function parameterized with  the real number $q$. 
 Because $c>0$ in the Gamma distribution function (Eq.\;\ref{eqgamma}), $q$ has to be larger than one in Eq.\;\ref{eq:Tsallis1}, but it  can be rewritten with the change of variable $q=2-q'$ in order to consider the cases where $q'<1$ \cite{wilk2012consequences, sebastian2011generalized}. 
 Thus, a generalized Boltzmann factor associated with the Gamma PDF  (Eq.\;\ref{eqgamma}) is defined as \cite{beck2003superstatistics, tsallis2019beyond, abe2001nonextensive}: 
\begin{equation}
B_q(E)= \exp_q({-\beta_0 E})
\end{equation}
 The  parameter $q$ can be thought of as a characteristic of the system's statistics, and is defined here by the ratio of standard variation and mean of the distribution $f_q(\beta)$  \cite{beck2003superstatistics}, noting that when $q=1$   there are no superposed fluctuations, and as it should be, the traditional exponential factor is recovered. Alternatively, the ordinary statistics are recovered in the limit $f_q(\beta)\to \delta(\beta - \beta_0)$ in Eq.\;\ref{eq:BE}.

\begin{figure}[t]
\begin{center}
{\includegraphics[]{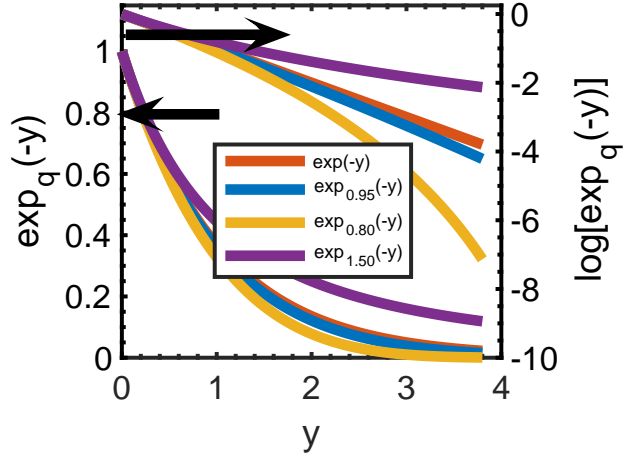}}
\caption{Plots of  $\exp(-y)$ and $\exp_q(-y)$ (Eq.\;\ref{eq:Tsallis1}) for different values of the parameter $q$ as a function of $y$}
\label{fig11}
\end{center}
\end{figure} 

We note some of the  properties of the  $q$-exponential function   \cite{abe2001nonextensive, beck2003superstatistics, beck2006stretched, allagui2021gouy}:  
\begin{itemize}
\item    for  $q<1$, $\exp_q(y)=0$ for $y<-1/(1-q)$ and $\exp_q(y)=\left[1+(1-q)y \right]^{1/(1-q)}$ for $y\geqslant -1/(1-q)$, 
  \item  for $q=1$, $\exp_q(y)=\exp(y)$ for $\forall y$, and 
  \item for $q>1$, $\exp_q(y)= \left[1+(1-q)y \right]^{1/(1-q)}$ for $y < 1/(q-1)$.  
\end{itemize} 
Fig.\;\ref{fig11}  shows  plots of the deformed $\exp_q(-y)$ as a function of $y$ for different values of the parameter $q$ \cite{borges2004possible, tsallis2019beyond}. The usual exponential function is also shown for reference. The logarithm of $\exp_q(-y)$ provided in the same figure shows linear relationship with $y$ only as $q\to 1$, otherwise, for $q<1$ the curve is convex, and for $q>1$ it is concave \cite{truhlar2001convex, silva2013uniform, aquilanti2017kinetics}. 
In addition to these algebraic properties, the function $\exp_q({-y})$ satisfies the anomalous, power-law rate equation $\mathrm{d}y/\mathrm{d}x = - y^q$ with $q \neq 1$ \cite{lyra1998nonextensivity, niven2006q}.

 Another important remark on the $q$-exponential and the $q$-Gaussian distributions is that they are the functions associated with some systems showing  quasi-stationary states, and   are the maximizing distributions for the non-additive Tsallis entropy given by \cite{tsallis1988possible, tsallis1999nonextensive,  tsallis2004should, abe2001nonextensive} :  
\begin{equation}
S_q 
= -k \sum_i p_i^q \ln_q(p_i)
\end{equation}
 where $k$ is a positive constant, $q\neq 1$  (also known as the entropic index), 
 and the quantities $p_i=p(E_i)$ represent the probabilities for the occurrence of the $i^{\text{th}}$ microstate and satisfy $\sum_{i} p_i=1$. The function:
 \begin{equation}
\ln_q(y) =  \frac{y^{(1-q)}-1}{1-q} \quad (y>0)
\end{equation}
 denotes the $q$-logarithm, inverse of the $q$-exponential i.e. $\ln_q[\exp_q(y)]=\exp_q[\ln_q(y)]=y$ \cite{tsallis2004should}. 
In this case, the underlying mathematical mechanism is now the generalized CLT \cite{umarov2008aq}.  
 It is clear that in the limiting case of $q\to 1$, $\ln_q(y)\to \ln(y) $, and one recovers the standard BG entropy 
$S_1 = -k_B \sum_{i} p_i \ln (p_i)$ 
 where $k=k_B$ is the Boltzmann constant \cite{tsallis1988possible}.

Finally, the generalized $q$-deformed BV model we propose, and incorporating the power-law distribution  given by the $q$-exponential function for describing the charge transfer reaction rates dependence on the overpotential and temperature, is  given by:
 \begin{align}
i^*_q &= \left[ 1- (1-q) (-\alpha \eta^*) \right]^{{1}/{(1-q)}} \nonumber \\
& - \left[ 1- (1-q) (1-\alpha)\eta^* \right]^{{1}/{(1-q)}} \\
&= \exp_q \left[-\alpha \eta^*\right] - \exp_q \left[ (1-\alpha)\eta^* \right]
\label{eqqBV}
\end{align}
We assumed that the parameter $q$ is the same for both half reactions. 
Again, recovering the ordinary expression of BV (Eq.\;\ref{eq10}) is obtained at the limit $q \to 1$. Furthermore, using the the expansion of the $q$-exponential function for sufficiently  small values of $y$, i.e. 
$ {\exp_q(-y)} = \exp(-y) \left[ 1+ \frac{1}{2}(q-1)y^2 - \frac{1}{3}(q-1)^2 y^3 + \ldots \right]
$, 
one can also recover the usual BV model  \cite{beck2003superstatistics}.

\subsection{$\kappa$-deformed BV model}

In the same way used for the formulation of the $q$-deformed BV model, we propose the following $\kappa$-deformed BV model:
 \begin{align}
i^*_{\kappa} 
= \exp_{\kappa} \left[-\alpha \eta^*\right] - \exp_{\kappa} \left[ (1-\alpha)\eta^* \right]
\label{eqkBV}
\end{align}
where the $\kappa$-deformed exponential function of $y$ is given by \cite{kaniadakis2001non}:
\begin{equation}
\exp_{\kappa}(y) \equiv \left(  \sqrt{1 +  \kappa^2 y^2} + \kappa y  \right)^{1/\kappa} = \exp \left[ \frac{1}{\kappa} \sinh^{-1} (\kappa y) \right]
\label{eqexpk}
\end{equation}
with $0\leqslant \kappa < 1$.   The function   $\exp_{\kappa}(y)$   emerges from a continuous linear combination of an infinity of standard exponentials as \cite{kaniadakis2013theoretical}:
\begin{equation}
\exp_{\kappa}(-\beta_0 E) = \int\limits_0^{\infty} \frac{1}{\kappa \beta} J_{1/\kappa} \left( \frac{\beta}{\kappa \beta_0} \right) \exp({-\beta E})  \mathrm{d}\beta 
\end{equation}
where $J_{\nu}(y)$ is the Bessel function of the first kind. This is equivalent to how the Gamma distribution is the weight function in the generalized Boltzmann factor given by Eq.\;\ref{eq:BE} that led  to the $q$-exponential function.   
Some of the basic properties of $\exp_{\kappa}(y)$ are \cite{kaniadakis2020kappa}:
\begin{itemize}
\item $\exp_{\kappa \to 0}(y) = \exp(y) $ and 
 $\exp_{\kappa}(y\to 0) = \exp(y) $ ($\exp_{\kappa}(0)=1$);
\item $\exp_{\kappa}(y) = \exp_{-\kappa}(y)$;
\item  for $y\to  \infty$, 
$\exp_{\kappa}(-y)$  is $ \sim (2\kappa y)^{-1/\kappa}  $ and  
$\exp_{\kappa}(y ) = +\infty$. 
\end{itemize}
Its associated inverse function is the $\kappa$-logarithm:
\begin{equation}
\ln_{\kappa}(y)= \frac{1}{\kappa} \sinh(\kappa \,\ln(y)) = \frac{y^{\kappa}  - y^{-\kappa}}{2 \kappa}
\end{equation}
  giving $\ln_{\kappa}(\exp_{\kappa} (y))  =   \exp_{\kappa}(\ln_{\kappa} (y)) = y$. 
  Kaniadakis' entropy associated with the $\kappa$-statistics is obtained by replacing the logarithm in the expression for the standard BG entropy by the $\kappa$-logarithm \cite{kaniadakis2001non, kaniadakis2013theoretical, abreu2018tsallis, kaniadakis2017composition}.

\begin{figure}[t]
\begin{center}
{\includegraphics[]{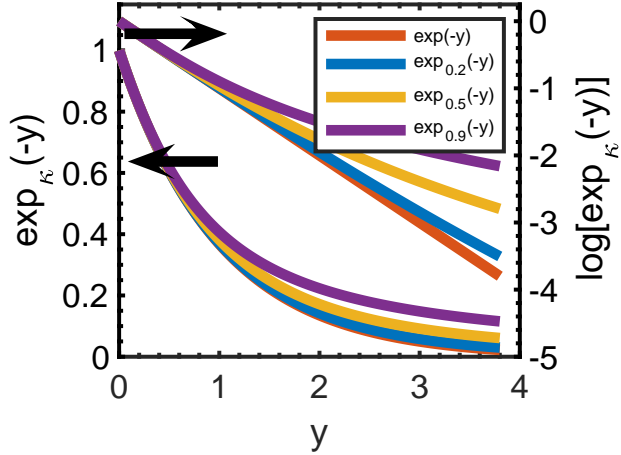}}
\caption{Plots of $\exp(-y)$ and $\exp_{\kappa}(-y)$ (Eq.\;\ref{eqexpk}) for different values of     $\kappa$ as a function of $y$}
\label{fig22}
\end{center}
\end{figure}

Plots  of  the standard  
$\exp(-y)$ vs. $\exp_{\kappa}(-y)$ for different values of the  parameter $\kappa$ are provided in Fig.\;\ref{fig22}.

\section{Results and discussion}
\label{simulation}

\begin{figure}[t]
\begin{center}
\includegraphics[]{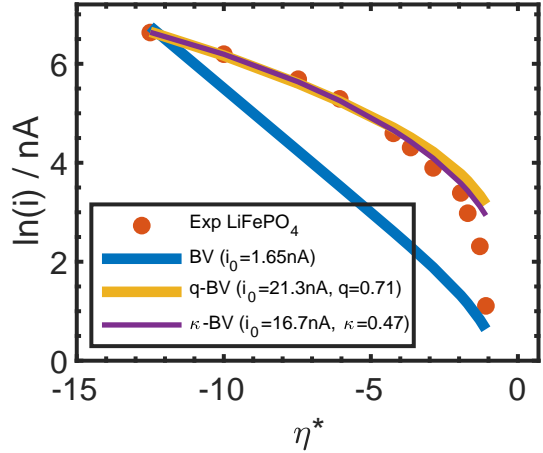}
\caption{Tafel plots of experimental data by Munakata et al.  \cite{munakata2012evaluation, bai2014charge} conducted on single   carbon-coated  \ce{LiFePO4} secondary particle (porous with 24\,$\mu$m diameter);  the  potential drop is originally measured with respect to \ce{Li/\ce{Li^+}} but is shown here in dimensionless form as done by Bai and Bazant \cite{bai2014charge}. We also show data fitting with standard BV model (Eq.\;\ref{eq10}), 
the $q$-deformed BV model (Eq.\;\ref{eqqBV}) and
 the $\kappa$-deformed BV model (Eq.\;\ref{eqkBV}) 
  with the symmetry factor $\alpha$ set to 0.5}
\label{fig3}
\end{center}
\end{figure}

In Fig.\;\ref{fig3} we show the  Tafel plots of electrochemical  measurements carried out  on 
 a single   carbon-coated  \ce{LiFePO4} particle as-is (i.e. in non-composite structure without interference from the effect of  binders and/or other additive conducting materials)   
 from Munakata et al.  \cite{munakata2012evaluation} and as adjusted by Bai and Bazant \cite{bai2014charge}  (avoiding concentration polarization effects).  Values of current in this single particle technique are usually very small which makes it acceptable to neglect the ohmic $iR$ drop. 
       Low-magnification  SEM of the target \ce{LiFePO4} particle (see Fig.\;2 in ref. \cite{munakata2012evaluation}) shows that it is spherical in shape, of about 24\,$\mu$m in diameter, and consisting of agglomerates of many 100 to 200\,nm-sized primary  particles with   inter-particle porosity and some defects. 
 The specific capacity of the   particle was estimated  to be 1.5\,nA\,h, and   the used discharge current was 750\,nA which took just 4\,s for full discharge, i.e. 900\,C  \cite{munakata2012evaluation}. 
From the figure, $\ln (i)$ vs. the (normalized)  voltage drop is clearly  curved and not as expected for traditional Tafel plots. 
This convex deviation from linearity  was attributed by Munakata et al.  \cite{munakata2012evaluation} to the distributions of electric  potential and current density, and also to the distribution of \ce{Li^+} concentration within the porous single particle electrode during charging and discharging, which is usually not observed when non-porous flat electrodes are considered. This nonlinear behavior becomes more significant when high rates are applied, which can be further explained as follows. Let us consider the discharge scheme consisting of the steps (i) \ce{Li^+} diffusion from the bulk electrolyte to the particle surface, followed by (ii) charge transfer at the particle/electrolyte interface, and then (iii) slow solid-phase diffusion   of  \ce{Li^+} or polaron diffusion from the surface to center of the particle coupled with phase transition from \ce{FePO4} to \ce{LiFePO4} \cite{munakata2012evaluation}. Step (iii) is the determining step if there is a large spatial gradient of \ce{Li^+} concentration within the particle from  surface to center which happens at high discharge current rates, whereas step (ii) may become the controlling step when low  rates are applied.  In other words, the system can be viewed as a combination of a spectrum of superposed kinetic processes of different origins on the electrode/electrolyte system. If the constituting subsystems are assumed temporarily in local equilibrium and follow Boltzmann exponential profiles  with different statistics, it can be well described with the modified BV models  as discussed above. 

In the same Fig.\;\ref{fig3}, we show the fitting curves performed with Eq.\;\ref{eqqBV} ($q$-BV model), Eq.\;\ref{eqkBV} ($\kappa$-BV model), and Eq.\;\ref{eq10} (standard BV model) with $\alpha=0.5$.  
We used MATLAB  R2019b \emph{lsqcurvefit} function for nonlinear curve-fitting in least-squares sense with the same fittings constraints and    tolerances for all models for fair comparison.  
When considering the traditional BV model, the data is poorly fitted with a straight line in the semilogarithmic plane of  $\ln (i)$ vs. $\eta^*$, noting that a better fit can be obtained if a smaller portion of the data is selected closer to $\eta^*=0$ (not shown here). 
 The goodness-of-fit using the normalized root mean square error (NRMSE) as the cost function is found to be 0.48, knowing that a value of 1 indicates a perfect fit to the data and $-\infty$ a bad one.  The BV model cannot be justified here given the morphological structure of the electrode as described in ref. \cite{munakata2012evaluation}. 
From the fitting with  the $q$-exponential and $\kappa$-exponential modified BV models, however, it is clear that the curved behavior of the data is closely captured. Convex or concave curvatures can be realized depending on the value of the parameter $q$ for the $q$-deformed model as shown in Fig.\;\ref{fig11}. Here we found the best fit to be with $q=0.71<1.00$, which phenomenologically indicates the extent of the system's departure from   BG statistics and the associated assumption of thermodynamic equilibrium. For the $\kappa$-deformed BV model we obtained $\kappa=0.47$, which can also be interpreted in a same way, i.e. the extent of the deviation or dispersion of the data from the usual exponential-based BV model that we can retrieve when $\kappa \to 0$.  
 The values of   the pre-factor current exchange densities are $i_0=21.3$\,nA and $i_0=16.7$\,nA for the $q$- and $\kappa$-deformed models, respectively. The NRMSE fitness values are 0.93    and        0.96, respectively, which are very close to  1. 
The   fittings by both   models are very close to each other given the power-law asymptotic behavior of both deformed  functions at the relatively large   values of the input overpotential. Thus,  we are not in the measure to promote or discriminate any of  the two  without enough information about the local statistics of the electrode. 

We note also that the fits are  in close agreement with Bai and Bazant's results \cite{bai2014charge}  based on   MHC theory. 
The MHC rate is, however, expressed in terms on indefinite integral of the exponential Boltzmann factor (which includes in its argument an extra term representing a reorganization energy) with  Fermi-Dirac statistics of electron energies distributed around the electrode potential. This makes the MHC model less attractive   from a computational point-of-view. On the other hand, both of the deformed BV models are clearly much simpler and easy to implement with comparable  accuracy and fitting capabilities.

\begin{figure}[t]
\begin{center}
\includegraphics[]{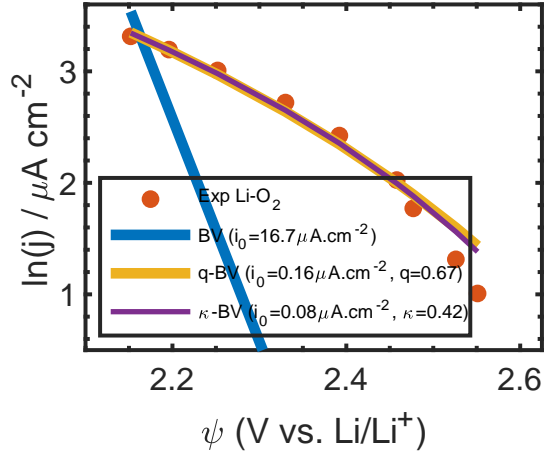}
\caption{Tafel plots of experimental data by Viswanathan et al. \cite{viswanathan2013li} for the discharge of \ce{Li}-\ce{O2} battery.  We also show data fitting with the 
standard  BV model (Eq.\;\ref{eqBVio}),
 $q$-deformed BV model (Eq.\;\ref{eqqBV}) and
  $\kappa$-BV model (Eq.\;\ref{eqkBV}) 
  with the symmetry factor $\alpha=0.5$}
\label{fig4}
\end{center}
\end{figure} 

The curvature of the Tafel plots for other materials and systems may be different, which can still be captured by the free parameters $q$ or $\kappa$ depending on the model as shown in Fig.\;\ref{fig4}. The figure depicts the experimental Tafel plots of extracted from Viswanathan et al. \cite{viswanathan2013li} for the discharge of \ce{Li}-air (or \ce{Li}-\ce{O2}) battery.  The net reaction is $\ce{2Li} + \ce{O2} \rightleftharpoons \ce{Li2O2}$ with the  battery discharge described by the forward direction. 
The nonlinear curve has been attributed to a complex crystal growth and dissolution mechanism of \ce{Li2O2} which can occur on different crystal facets or terminations on the electrode, on different sites (terrace, step or kink), and could involve different combinations of nucleation and diffusion  \cite{viswanathan2013li} or other mechanisms  
\cite{sankarasubramanian2017elucidating}.  
From the figure, it is clear that the standard BV model also fails to adequately fit these types of data, whereas based on different statistics than BG theory, the $q$- and $\kappa$-deformed BV models successfully followed the trend of the curve with ($q=0.67$, $i_0=0.16\,\mu\text{A\,cm}^{-2}$) and ($\kappa=0.42$, $i_0=0.08\,\mu\text{A\,cm}^{-2}$) respectively. The goodness-of-fit between the models and the experimental data are -0.11, 0.89 and 0.91, respectively.

\section{Conclusion}
\label{conclusion}

In this contribution, we proposed and verified the application of two  deformed versions of the BV model for the analysis of convex semilogarithmic current-overpotential profiles observed with  some battery data. The models assume that the nonequilibrium electrode/electrolyte system can be viewed as a multitude of subsystems temporarily in local equilibrium, and thus follow the Boltzmann exponential trend but   with different statistics. These fluctuations are taken to be inverse temperature fluctuations that can be  correlated for instance to spatial inhomogeneities of the interface geometry, long-range interactions, particle trapping and partial charge transfer, as well as distribution of thermodynamic quantities. Applied to two different examples of battery data \cite{munakata2012evaluation, viswanathan2013li, bai2014charge}, both  deformed models with only one free parameter each (based on the $q$- or $\kappa$-exponential functions) showed very close agreement with the experiments. The extra degree of freedom in the modified BV models is related to the extent of deviation of the data from  BG statistics assumed in classical BV, which itself can be retrieved when $q\to 1$ in Eq.\;\ref{eqqBV} or $\kappa \to 0$ in Eq.\;\ref{eqkBV}. The deformed BV models can in principle be applied to other types of usual or unusual reaction data   and power-law relaxation behavior found in corrosion reactions, sensors,  electrocatalytic processes, solar cells, etc.

%
%

\section*{Acknowledgement}
   
This work is supported by NSF project \#2126190 (C.W \& A.A.)


%


 \providecommand{\latin}[1]{#1}
\makeatletter
\providecommand{\doi}
  {\begingroup\let\do\@makeother\dospecials
  \catcode`\{=1 \catcode`\}=2 \doi@aux}
\providecommand{\doi@aux}[1]{\endgroup\texttt{#1}}
\makeatother
\providecommand*\mcitethebibliography{\thebibliography}
\csname @ifundefined\endcsname{endmcitethebibliography}
  {\let\endmcitethebibliography\endthebibliography}{}

%
%
%
%
%

 \end{document}